# Joint Adaptive Modulation Coding and Cooperative ARQ over Relay Channels-Applications to Land Mobile Satellite Communications[†]

Morteza Mardani*, Jalil S. Harsini*, Farshad Lahouti*, Behrouz Eliasi**

*Wireless Multimedia Communication Lab., School of E&CE, University of Tehran
**Iran Telecommunication Research Center, PO Box 14155-3961, Tehran 14399, Iran
Emails: m.mardani@ece.ut.ac.ir, j.harsini@ece.ut.ac.ir, lahouti@ut.ac.ir, eliasi@itrc.ac.ir

**Summary**

In a cooperative relay network, a relay node (R) facilitates data transmission to the destination node (D), when the latter is unable to decode the source node (S) data correctly. This paper considers such a system model and presents a cross-layer approach to jointly design adaptive modulation and coding (AMC) at the physical layer and cooperative truncated automatic repeat request (ARQ) protocol at the data link layer. We first derive a closed form expression for the spectral efficiency of the joint cooperative ARQ-AMC scheme. Aiming at maximizing this performance measure, we then optimize two AMC schemes for S-D and R-D links, which directly satisfy a prescribed packet loss rate constraint. As an interesting application, we also consider the problem of joint link adaptation and blockage mitigation in land mobile satellite communications (LMSC). We also present a new relay-assisted transmission protocol for LMSC, which delivers the source data to the destination via the relaying link, when the S-D channel is in outage. Numerical results indicate that the proposed schemes noticeably enhances the spectral efficiency compared to a system, which uses a conventional ARQ-AMC scheme at the S-D link, or a system which employs an optimized fixed rate cooperative-ARQ protocol.

## 1. Introduction

Recently, cooperative communication has attracted a lot of research attention as a promising technique to achieve diversity gain in wireless networks. In particular, cooperative automatic repeat request (ARQ) is a link-level protocol which exploits the spatial diversity of the relay channel. It outperforms a conventional ARQ scheme when the source to destination channel is subject to a high temporal correlation [8], [10]. The main idea behind this protocol is to jointly exploit the benefits of two relaying protocols: (i) the incremental decode and forward relaying protocol, which prescribes a retransmission via relay only when the destination decodes the source data in error [6], and (ii) the selection decode and forward protocol, which verifies that the data is received correctly at the relay, prior to a possible relay to destination retransmission [6].

[†]This work has been supported in part by the Iran Telecommunications Research Center, and has been accepted for presentation in parts at the IEEE International Symposium on Telecommunications, Tehran, Iran, August 2008, and IEEE International Symposium on Wireless Communication Systems, Reykjavik, Iceland, October 2008.

However, time varying nature of wireless links limits the communication performance over these systems for provisioning of stringent quality of service (QoS) requirements. Adaptive Modulation and Coding (AMC) is known as a powerful technique to enhance the system spectral efficiency for communications over wireless fading channels [1], [2]. Thanks to high spectral efficiency, AMC schemes are already included in wireless communication standards such as HIPERLAN/2, IEEE 802.11 and IEEE 802.16e. It is also of great interest in satellite communications and has been adopted in the DVB-S2 standard [3]-[5].

There are several research studies on the topic of cooperative ARQ in the relay channel, e.g., [7]-[12]. Utilizing a distributed space-time coded retransmission protocol, in [7] a truncated cooperative-ARQ is proposed, which exploits adaptive cooperative diversity, where the relay nodes are selected using a cyclic redundancy check (CRC) code. In [8], for several cooperative ARQ protocols, the link layer performance including both throughput and packet loss rate, is studied over slow fading channels. Their results illustrate that the cooperative ARQ protocol compared to the conventional ARQ, achieves better performance if the average SNR of the relay to destination (R-D) channel is better than a given threshold. In [9], closed form expressions for the throughput performance of several relay-based retransmission protocols, corresponding to different levels of cooperation, are presented. Based on the channel statistics (average SNR), they used a simulation setup to investigate the effect of optimized rate selection on the system throughput (no AMC is employed). In [10], a stop and wait cooperative ARQ protocol is developed and analyzed, for improved throughput and packet delay performance, over time-correlated fading channels. Delay performance of a set of cooperative ARQ protocols is also investigated in [11], where data frame arrival at the source node is modeled by a Poisson process. The research presented in [12], verifies that utilizing a cooperative retransmission strategy compared to a conventional ARQ, reduces both system load and packet delay in mobile satellite communication systems suffering from channel blockage effects.

The idea of applying AMC to a wireless relay network is investigated in [14] and [15]. In [14] for a cooperative decode and forward relay network, it is demonstrated that both power and rate adaptation at the source and relay nodes, lead to an improved network throughput compared to a direct transmission system. In [15], an OFDM based wireless relay network is considered and aiming at optimizing the end-to-end instantaneous throughput, a joint relaying scheme (without ARQ) and AMC mode selection algorithm is proposed and solutions are provided in the form of lookup tables.

As mentioned above, in the literature, both the cooperative ARQ protocol and AMC over the relay channel are separately investigated. Nevertheless, the problem of designing discrete-rate AMC schemes in conjunction with a cooperative ARQ protocol in wireless relay networks has not been addressed so far. This will be promising especially in land mobile satellite communication (LMSC) systems, where the communication channel experiences both blockage and multipath fading effects.

The main contribution of this paper is to quantify the potential spectral efficiency gain achieved by the joint design of discrete-rate AMC with cooperative ARQ, while satisfying the QoS constraints of higher layers. To this end, we take a cross-layer design approach. We first derive an exact closed form expression for the spectral efficiency of joint cooperative ARQ-AMC scheme over block fading channels, when the number of retransmission attempts per packet at the relay node is finite. Then aiming at maximizing this performance measure, we propose an AMC-based rate adaptation policy for the relay channel, which guarantees a prescribed average packet loss rate (PLR) constraint. In the case where only the channel statistics are available, we also present an optimized rate selection policy for both transmission rates on the source to destination (S-D) and R-D links, which guarantees the required PLR. As an interesting application of the proposed scheme, we consider AMC design and blockage mitigation for land mobile satellite communications in presence of a relay.

Numerical results for both terrestrial links with Rayleigh fading and LMSC links show that the proposed cross-layer design for joint cooperative ARQ-AMC scheme achieves considerable spectral efficiency gain. In particular, it outperforms a joint conventional ARQ-AMC scheme designed for direct S-D link, an optimized fixed rate cooperative ARQ when different optimized transmission rates at the source and relay nodes are chosen based on the channel statistics, and AMC alone at the physical layer. Also, the results on LMSC system, demonstrates that if the relay retransmits source data to the destination node, when the S-D channel is in outage, an even higher spectral efficiency gain is achieved.

As a side result on the conventional ARQ-AMC scheme, we also observe that imposing different optimized target packet error rates (PER) on the transmission and the possible retransmissions of a packet leads to higher spectral efficiency compared to a scheme that considers identical target PERs on all transmissions, as presented in [16].

The rest of this paper is organized as follows. Section 2 describes the system and channel model. In Section 3, we first derive a closed-form expression for the spectral efficiency of an adaptive rate cooperative-ARQ scheme. We then present a cross-layer approach aiming at maximizing the spectral efficiency subject to packet-level QoS constraints. We also describe the cooperative ARQ in a fixed rate scenario. In section 4, we consider the application of the proposed scheme for LMSC. Numerical results are provided in section 5, while the concluding remarks are presented in sections 6.

## 2. System Model

### 2.1. System Description

As illustrated in Fig.1, we consider an adaptive rate wireless network composed of a S node, a R node and a D node , where each node is equipped with a single antenna. At the S node input packets from higher layers of stack are stored in the transmit buffer, grouped into frames, and then transmitted over the wireless channel on a frame by frame basis. We adopt the packet and frame structure as in [16], where each frame contains multiple packets based on the employed AMC mode, and each packet includes a CRC code for error detection. We assume a time-division-duplex (TDD) system for nodes and that each node does not transmit and receive simultaneously. Adopting a decode and forward strategy for the R node, the proposed cooperative-retransmission system operates as follows: First, the S node broadcasts a data frame to both D and R nodes, and they listen. Upon reception of a data frame by the D node, it checks the CRC for each packet separately, and transmits either a positive or negative acknowledgement (ACK or NACK). If the relay receives a NACK message from the destination, and it has successfully decoded the corresponding packet, it then retransmits this packet to the destination until it is received correctly, or a maximum allowable number of retransmissions is reached. Otherwise, the node S transmits a new data frame and the above procedure is repeated.

In the proposed analysis, it is assumed that the transmit buffer at the source node is always loaded with packets as in [16]. Therefore, we only consider the effect of channel processes on the system performance.

### 2.2. Channel Models and AMC Modes

In Fig. 1, we assume that both S-D and R-D wireless links are modeled as flat-fading channels with AWGN and stationary channel gains $\sqrt{h_{sd}}$ and $\sqrt{h_{rd}}$, respectively, while the S-R link is assumed to be an AWGN channel as in [12], and [13]. The latter assumption is valid, e.g., in a setting with fixed source and relay positions and a strong S-R link with a direct line of sight. We adopt a block fading model so that the channel gains remain constant over a frame period and vary from one frame to another independently [17]. Let $N_0$ be the one sided noise power spectral density and $W$ denote the system spectral bandwidth. In our analysis, we assume that both S and R nodes have the same constant transmit power level denoted by $\bar{P}$ . As a result, the instantaneous received SNR at the destination for the S-D and R-D channels are $\gamma_1 = \bar{P} h_{sd}/N_0 W$ and $\gamma_2 = \bar{P} h_{rd}/N_0 W$, respectively.

At the physical layer, AMC is employed for both S-D and R-D links based on their corresponding channel state information (CSI). We assume that perfect CSI is available at the destination and that the selected AMC modes are fedback to the source and relay nodes reliably and without delay. To employ the

AMC, the entire SNR range of S-D and R-D links are divided into *N+1* and *M+1* non-overlapping consecutive intervals, respectively. When the S-D channel SNR $\gamma_1$ falls in the interval $[\gamma_{1,n}, \gamma_{1,n+1})$, $n = 0,1,\dots,N$ where $\gamma_{1,0} = 0$ and $\gamma_{1,N+1} = \infty$, the mode *n* of AMC is chosen and the source transmits with rate $R_n^1$ from the rate set $\mathcal{R}^1 = \{R_n^1\}_{n=0}^N$. Also, when the R-D channel SNR $\gamma_2$ is in the interval $[\gamma_{2,m}, \gamma_{2,m+1})$, $m = 0,1,2,\dots,M$ where $\gamma_{2,0} = 0$ and $\gamma_{2,M+1} = \infty$, the relay transmits with rate $R_m^2$ from the rate set $\mathcal{R}^2 = \{R_m^2\}_{m=0}^M$. No signal is transmitted when the mode index *n=0* (*m=0*) is selected, corresponding to the link outage mode, i.e., $R_0^1 = 0$ ($R_0^2 = 0$). In the following, without loss of generality, we choose the same rate set for both source and relay nodes, i.e., $\mathcal{R}^1 = \mathcal{R}^2 = \mathcal{R} = \{R_n\}_{n=0}^N$.

In order to simplify the analysis, we approximate the PER for the AMC mode *n*, using the following expression [16]

$$PER_n(\gamma) \approx \begin{cases} 1, & \gamma < \Gamma_n \\ a_n \exp(-g_n \gamma), & \gamma \geq \Gamma_n \end{cases} \quad (1)$$

where the parameters $\{a_n, g_n, \Gamma_n\}$ are determined by curve fitting to the exact PER of mode *n*. This model is verified in [16].

## 3. Joint Design of Cooperative ARQ and AMC

In this section, we develop a cross-layer approach to jointly design AMC at the physical layer and cooperative ARQ at the data link layer, when the following QoS constraints are imposed by the packet service.

C1) Delay constraint: The maximum number of retransmission attempts per packet by the R node is limited to $N_r$. Accordingly, if a packet is not received correctly after the relay retransmissions, it is considered lost.

C2) PLR QoS constraint: At the data-link layer, the packet loss probability following $N_r$ possible relay retransmissions is to be less than a target PLR $P_{loss}$.

To this end, we first derive an exact closed form expression for the spectral efficiency performance of a joint cooperative ARQ-AMC scheme with a maximum of $N_r$ possible retransmissions; we then optimize this performance measure subject to a PLR constraint described in C2.

### 3.1. Spectral Efficiency

In [2], the spectral efficiency for an adaptive rate scheme is defined as the average number of information bits transmitted per symbol. Here, we develop a similar definition for the spectral efficiency of the proposed joint cooperative ARQ-AMC scheme.

*Proposition 1*: For the considered adaptive rate cooperative ARQ protocol, when the channel gains for transmission and retransmissions of a packet are independent, the average spectral efficiency is given by

$$\eta = \sum_{n=1}^{N} \frac{(1-(1-\varepsilon_n)\overline{PER}_n^{sd})}{L(n,\underline{m}^{(0)})} P_n^{sd} + \sum_{l=1}^{N_r-1} \sum_{n=1}^{N} \sum_{m_1=1}^{N} \cdots \sum_{m_l=1}^{N} \frac{(1-\varepsilon_n)}{L(n,\underline{m}^{(l)})}$$

$$\times \overline{PER}_n^{sd} P_n^{sd} \prod_{k=1}^{l-1} \overline{PER}_{m_k}^{rd} P_{m_k}^{rd} (1 - \overline{PER}_{m_l}^{rd}) P_{m_l}^{rd}$$

$$+ \sum_{n=1}^{N} \sum_{m_1=1}^{N} \cdots \sum_{m_{N_r}=1}^{N} \frac{(1-\varepsilon_n)\overline{PER}_n^{sd} P_n^{sd} P_{m_{N_r}}^{rd}}{L(n,\underline{m}^{(N_r)})} \prod_{k=1}^{N_r-1} \overline{PER}_{m_k}^{rd} P_{m_k}^{rd} \quad (2)$$

where $\gamma_{sr}$ is the S-R channel SNR, $\varepsilon_n = PER_n(\gamma_{sr})$ is the PER of S-R channel in mode $n$, $L(n,\underline{m}^{(l)}) = 1/R_n + \sum_{k=1}^{l} 1/R_{m_k}$, $\underline{m}^{(l)} = (m_1, m_2, \ldots, m_l)$, $P_n^{sd} = \int_{\gamma_{1,n}}^{\gamma_{1,n+1}} p_{\gamma_1}(\gamma) d\gamma$, $P_{m_k}^{rd} = \int_{\gamma_{2,m_k}}^{\gamma_{2,m_k+1}} p_{\gamma_2}(\gamma) d\gamma$ and

$$\overline{PER}_n^{sd} = \frac{1}{P_n^{sd}} \int_{\gamma_{1,n}}^{\gamma_{1,n+1}} PER_n(\gamma) p_{\gamma_1}(\gamma) d\gamma \quad (3)$$

$$\overline{PER}_{m_k}^{rd} = \frac{1}{P_{m_k}^{rd}} \int_{\gamma_{2,m_k}}^{\gamma_{2,m_k+1}} PER_{m_k}(\gamma) p_{\gamma_2}(\gamma) d\gamma, \quad k = 1,2,\ldots,N_r \quad (4)$$

*Proof*: The proof is provided in Appendix A.

For the special case of noiseless S-R channel, i.e., $\varepsilon_n = 0$, and the identical positions of the S and R nodes, the system reduces to a joint conventional ARQ-AMC scheme. In this case, the S-D and R-D channels have identical statistics and the following corollary describes the performance of this scheme.

*Corollary 1*: For an adaptive rate conventional ARQ scheme with a maximum number of retransmissions per packet $N_r$, the average spectral efficiency is obtained as follows

$$\eta = \sum_{l=1}^{N_r} \sum_{n_1=1}^{N} \cdots \sum_{n_l=1}^{N} \frac{(1-\overline{PER}_{n_l}^{sd}) P_{n_l}^{sd} \prod_{k=1}^{l-1} \overline{PER}_{n_k}^{sd} P_{n_k}^{sd}}{L(\underline{n}^{(l)})} + \sum_{n_1=1}^{N} \cdots \sum_{n_{N_r+1}=1}^{N} \frac{P_{n_{N_r+1}}^{sd} \prod_{k=1}^{N_r} \overline{PER}_{n_k}^{sd} P_{n_k}^{sd}}{L(\underline{n}^{(N_r+1)})} \quad (5)$$

where $L(\underline{n}^{(l)}) = \sum_{k=1}^{l} 1/R_{n_k}$ and $\underline{n}^{(l)} = (n_1, n_2, \ldots, n_l)$.

*Proof*: The proof is straightforward from Proposition 1 by substituting $\varepsilon_n = 0$ and $p_{\gamma_1}(\gamma) = p_{\gamma_2}(\gamma)$.

### 3.2. Optimizing the Spectral Efficiency

Based on the performance metric derived above, here we propose a cross-layer design for adaptive rate cooperative-ARQ system in Fig. 1, which maximizes the system spectral efficiency subject to a PLR constraint. The desired optimization problem can be formulated as follows

$$\max_{\{\gamma_{1,n},\gamma_{2,m}\}_{n,m=1}^{N}} \eta \quad \text{subject to}$$
$$C: \overline{PLR} \leq P_{loss} \quad (6)$$

where $\overline{PLR}$ is the system average PLR, and the constraint C states that the average PLR is not greater than the target PLR as described in C2. In the followings, we present an analysis for the case of $N_r = 1$. It is noteworthy that the proposed analysis can be easily extended to the case of $N_r > 1$. Although in general, employing a larger $N_r$ may lead to a smaller achievable PLR. However, as demonstrated in [18] for the case of a point-to-point block-fading wireless link AMC-ARQ, in the practical range of the target PLR in C2, $N_r = 1$ almost achieves the maximum possible spectral efficiency gain for an transmission scheme over channels. Using (2), the average system spectral efficiency for $N_r = 1$ is given by

$$\eta = \sum_{n=1}^{N} R_n (1 - (1 - \varepsilon_n)\overline{PER}_n^{sd})P_n^{sd} + \sum_{n=1}^{N}\sum_{m=1}^{N} \frac{R_n R_m}{R_n + R_m}(1 - \varepsilon_n)\overline{PER}_n^{sd} P_m^{rd} P_n^{sd} \quad (7)$$

In order to solve the problem in (6), we first develop the following Proposition.

*Proposition 2:* The average PLR of the considered adaptive rate cooperative ARQ protocol for $N_r = 1$ is given by

$$\overline{PLR} = \left(\frac{\sum_{n=1}^{N} \overline{PER}_n^{sd} P_n^{sd}}{\sum_{n=1}^{N} P_n^{sd}}\right)\left(\frac{\sum_{m=1}^{N} \overline{PER}_m^{rd} P_m^{rd}}{\sum_{m=1}^{N} P_m^{rd}}\right) + \left(\frac{\sum_{n=1}^{N} \varepsilon_n \overline{PER}_n^{sd} P_n^{sd}}{\sum_{n=1}^{N} P_n^{sd}}\right) \times \left(1 - \frac{\sum_{m=1}^{N} \overline{PER}_m^{rd} P_m^{rd}}{\sum_{m=1}^{N} P_m^{rd}}\right) \quad (8)$$

where $P_n^{sd}$, and $\overline{PER}_n^{sd}$, denote the probability that the mode $n$ of AMC is chosen, and the average PER of mode $n$, respectively for the S-D link. The parameters, $P_m^{rd}$ and $\overline{PER}_m^{rd}$, are similarly defined for the R-D link.

*Proof:* The proof is provided in appendix B.

*Corollary 2:* The average PLR of the adaptive rate conventional ARQ protocol is obtained as follows

$$\overline{PLR} = \left(\frac{\sum_{n=1}^{N} \overline{PER}_n^{sd} P_n^{sd}}{\sum_{n=1}^{N} P_n^{sd}}\right)\left(\frac{\sum_{n=1}^{N} \overline{PER}_n^{sd,r} P_n^{sd,r}}{\sum_{n=1}^{N} P_n^{sd,r}}\right) \quad (9)$$

where the superscript $r$ refers to the retransmission parameters.

*Proof:* The proof is straightforward from Proposition 2 by substituting $\varepsilon_n = 0$ and $p_{\gamma_1}(\gamma) = p_{\gamma_2}(\gamma)$. The motivation and implication of distinctive transmission and retransmission parameters in this setting is elaborated in the followings and in section 5.1.

Using Proposition 2, in the following we propose an approach to convert the total PLR constraint C in (6) into two separate PER constraints over S-D and R-D links, so that the AMC design process over each of these links can be solved separately.

In this formulation, we design adaptive schemes over the S-D and R-D links, in order to achieve two different target average PERs, $P_{t,sd}$ and $P_{t,rd}$, i.e.,

$$\overline{PER}_n^{sd} = P_{t,sd}, \quad n = 1,2,...,N \quad (10)$$

and

$$\overline{PER}_m^{rd} = P_{t,rd}, \quad m = 1,2,...,N \quad (11)$$

Inserting the $\overline{PER}_n^{sd}$ and $\overline{PER}_m^{rd}$ from (10) and (11), into the PLR constraint C in (8), and equating it with the target PLR $P_{loss}$, we obtain the following relation between the target PLR and PERs in the system

$$P_{loss} = P_{t,sd}P_{t,rd} + \bar{\varepsilon}P_{t,sd}(1 - P_{t,rd}) \tag{12}$$

where,

$$\bar{\varepsilon} = \frac{\sum_{n=1}^{N} \varepsilon_n P_n^{sd}}{\sum_{n=1}^{N} P_n^{sd}} \tag{13}$$

The design problem is to find the optimal target PERs, $P_{t,sd}^*$ and $P_{t,rd}^*$, such that the system spectral efficiency is maximized, while satisfying the equation (12). The following algorithm describes a search method for this purpose.

Step 1) Choose $P_{t,sd} \in \mathcal{P}$, where the set $\mathcal{P}$ is

$$\mathcal{P} = \{P_{t,sd}: P_{loss} < P_{t,sd} < 1\}$$

Step 2) Design AMC for the S-D link based on the given $P_{t,sd}$, and equation (10), following the approach suggested in [19] and Remark 1 below.

Step 3) Compute $\bar{\varepsilon}$ using equation (13).

Step 4) Given $P_{loss}$, $\bar{\varepsilon}$, $P_{t,sd}$, using (12), we obtain

$$P_{t,rd} = \frac{P_{loss} - \bar{\varepsilon}P_{t,sd}}{P_{t,sd}(1 - \bar{\varepsilon})} \tag{14}$$

Step 5) Design AMC for the R-D link based on the given $P_{t,rd}$, and equation (11), following the approach suggested in [19] and Remark 1 below.

Step 6) Compute $\eta(P_{t,sd})$ using (7).

Step 7) Repeating steps 1 to 6, determine the optimal $P_{t,sd}$ as follows

$$P_{t,sd}^* = \underset{P_{t,sd} \in \mathcal{P}}{\mathrm{argmax}}\ \eta(P_{t,sd}) \tag{15}$$

Once, $P_{t,sd}^*$ and subsequently $P_{t,rd}^*$ are obtained, the design process is completed. A special case of interest is to consider a S-R channel with high SNR ($\bar{\varepsilon} \approx 0$). In this case the system target PLR is divided between S-D and R-D links such that $P_{t,rd}^* P_{t,sd}^* = P_{loss}$.

Since the objective function of the optimization problem in (15) is a complicated function of the target PER $P_{t,sd}$, in order to solve it, one may devise more efficient search algorithms. Specifically, in a similar case in [20], a low complexity gradient-based search method is presented.

**Remark 1:** For a single wireless link such as the S-D channel, the AMC design procedure in [19] is based on satisfying the equation (10) for each of the transmission modes (This is also true for the R-D link). However, our experiments show that for large values of the target PER $P_{t,sd}$, these equations may not be met with equality over distinct S-D and R-D links. This observation affects the constraint C in problem (6) in a way that the achievable system PLR will be smaller than $P_{loss}$. In this case, the set $\mathcal{P}$ in step 1 can be reduced to $\mathcal{P} = \{P_{t,sd}: P_{loss} < P_{t,sd} < P_{t,sd}^{up}\}$, where $P_{t,sd}^{up} = \max_n\{P_{t,n,sd}^{up}\}$ and $P_{t,n,sd}^{up}$ is specified as follows

$$P_{t,n,sd}^{up} = \frac{\int_{\Gamma_n}^{\Gamma_{n+1}} a_n \exp(-g_n\gamma) p_{\gamma_1}(\gamma) d\gamma}{\int_{\Gamma_n}^{\Gamma_{n+1}} p_{\gamma_1}(\gamma) d\gamma}, n = 1,2,\ldots,N \tag{16}$$

The upper bounds for the mode average PERs in (16) are derived based on the SNR lower bounds for mode switching levels, i.e., $\gamma_n = \Gamma_n, \forall n$.

### 3.3. Fixed rate cooperative-ARQ Scheme

In general, rate adaptation in a wireless relay network as depicted in Fig. 1, requires the channel CSI for both the S-D and R-D links. In some scenarios providing instantaneous (per frame) CSI may not be feasible. In such cases, our design can be modified to obtain the optimized fixed transmission rates for S-D and R-D links provided that the channel statistics of these channels are available at the S and R nodes, respectively.

Let us consider the problem of optimized rate pair $(R_n, R_m)$ selection for the source and relay nodes, based on the following optimization problem,

$$\max_{n,m} \eta(n,m) \text{ subject to} \tag{17}$$
$$C: \overline{PLR}(n,m) \leq P_{loss}$$

in which, following the same procedure as presented in Appendices A and B, the average spectral efficiency and the average PLR for a fixed rate cooperative-ARQ scheme can be obtained as follows

$$\eta(n,m) = R_n\left(1 - (1-\varepsilon_n)\frac{R_n}{R_n+R_m}\overline{PER}_{sd}(n)\right) \tag{18}$$

$$\overline{PLR}(n,m) = \overline{PER}_{sd}(n)\overline{PER}_{rd}(m) + \varepsilon_n\overline{PER}_{sd}(n)(1 - \overline{PER}_{rd}(m))$$

where, based on equation (1), we have

$$\overline{PER}_{sd}(n) = \int_0^\infty PER_n(\gamma)p_{\gamma_1}(\gamma)d\gamma$$
$$= \int_0^{\Gamma_n} p_{\gamma_1}(\gamma)d\gamma + \int_{\Gamma_n}^\infty a_n \exp(-g_n\gamma) p_{\gamma_1}(\gamma)d\gamma \tag{19}$$

The value of $\overline{PER}_{rd}(m)$ may be obtained similarly using the R-D channel parameters.

**Remark 2:** In a fixed rate scenario, the average PLR constraint C in (17) may not be satisfied for the entire range of $(\bar{\gamma}_1, \bar{\gamma}_2)$. In fact, given a rate pair $(R_n, R_m)$, our experiments show that there is a power threshold $\bar{P}_{th}$, for which the PLR constraint C is satisfied only when, $\bar{P} > \bar{P}_{th}$. As a result, for $\bar{P} < \bar{P}_{th}$ the spectral efficiency is zero.

In order to solve the problem in (17), an iterative procedure similar to the one presented in Section 3.2 for the case of adaptive rate cooperative ARQ, can be devised in a straight forward manner.

## 4. Applications to Blockage Mitigation in Land Mobile Satellite Links

In this section, we consider the application of the proposed joint cooperative ARQ-AMC scheme in land mobile satellite communications. The aim is to facilitate efficient communications in presence of satellite channel variations and blockage.

Recently, utilizing the AMC in the LMSC systems is well motivated by the advances in channel estimation and predication techniques for tracking of time varying satellite channels [4]. On the other hand, satellite to mobile links suffer from channel blockages, which appear as deep fades over long periods of time [27]. As an error-control mechanism, conventional ARQ protocol is used to combat the burst errors of such channels [21], [22]. However, this in turn increases the satellite load and the overall latency in the system, especially given the highly correlated nature of such channels and potentially large number of required retransmissions [12]. As validated in [12] and [23], cooperative relaying appears as a promising technique to mitigate the channel blockage and to extend the satellite coverage, at the expense of involving a relay terminal for packet retransmissions. To apply the proposed cross-layer design to LMSC, we first introduce the LMSC system and channel model. Then, for this specific application, we present system design considerations to efficiently mitigate the S-D channel outage with the aid of the relay terminal.

### 4.1. LMSC System and Channel Model

We consider the downlink of a packet based geosynchronous satellite system assisted with a relay terminal. In this system, the satellite acts as the source node, the relay node can be an airborne node, i.e., a high altitude platform station [24], [25] or a satellite ground terminal, e.g., a gap filler [26], and the destination node is a mobile terminal. As in [12], we model each of the S-D and R-D channels by a two-state Markov blockage channel, where the states correspond to the unblocked and blockage modes. The satellite-relay channel is also considered as a high SNR AWGN channel.

In the unblocked channel state, the channel gain amplitude for both of the S-D and the R-D links, follow a Rician distribution due to the presence of a line of sight (LOS) path. As a result, the corresponding channel SNR $\gamma$ has a Chi-square distribution with the following probability density function (PDF)

$$p_{Rice}(\gamma) = \frac{(1+k)e^{-k}}{\bar{\gamma}_u} \exp\left(\frac{-(1+k)\gamma}{\bar{\gamma}_u}\right) I_0\left(2\sqrt{\frac{k(k+1)\gamma}{\bar{\gamma}_u}}\right) \qquad (20)$$

where $I_0(\cdot)$ is the modified Bessel function of order zero, and the parameters $k$ and $\bar{\gamma}_u$ denote the Rice factor and the average SNR, respectively. In the blockage channel state, due to the shadowing effect and lack of a LOS, the mean received signal power follows a Lognormal distribution, and the amplitude of multipath fading obeys a Rayleigh distribution. As a consequence the channel SNR $\gamma$ follows the following PDF [27]

$$p_{Ray/Log}(\gamma) = \int_0^\infty \frac{e^{-\gamma/w}}{w} \frac{\xi}{\sqrt{2\pi}\sigma_s w} \exp\left\{\frac{-(10\log_{10}^w - \mu_s)^2}{2\sigma_s^2}\right\} dw \tag{21}$$

where $\xi = 10/\ln 10$. The parameters, $\mu_s$ and $\sigma_s$ denote the mean and standard deviation of the channel SNR in blockage state, respectively.

According to the above discussion, each realization of the channel SNR $\gamma$, is governed by the Lutz distribution as follows [27]

$$p_\gamma(\gamma) = (1-A)p_{Rice}(\gamma) + Ap_{Ray/Log}(\gamma) \tag{22}$$

where $A$ is the blockage state probability. Obviously, when the S-D channel experiences an outage, a direct reliable communication is not possible. In such cases, since the relay terminal may be able to communicate with the source node reliably, we use a relay-assisted transmission protocol for LMSC system, which allows source data transmission via the relay link. The main idea here is that the relay node is positioned such that the outage probability for the corresponding R-D link is smaller compared to the direct S-D link (see Table I). As elaborated below, in the proposed scheme, a new transmission mode is added to the AMC design for the S-D link, assuming that the source may transmit data in the outage mode $\gamma_1 \in [\gamma_{1,0}, \gamma_{1,1})$ with rate $R_0 > 0$, via the relay.

### 4.2. Joint AMC-Cooperative ARQ Scheme for LMSC

There are two main differences between the LMSC system model considered, and that presented in section 2 for a terrestrial channel model. First, the relay node in LMSC system is assumed to receive the source data reliably, i.e., $\bar{\varepsilon} = 0$ (see eq. (13)), and the second difference is that the source in this system can transmit data in S-D outage mode with rate $R_0$. Based on these assumptions, the following corollary presents an expression for the spectral efficiency of the joint cooperative ARQ-AMC scheme in LMSC system.

*Corollary 3:* The average spectral efficiency of the considered LMSC system for $N_r = 1$, is given by

$$\eta = \sum_{n=0}^{N} R_n \left(1 - \overline{PER}_n^{sd}\right) P_n^{sd} + \sum_{n=0}^{N} \sum_{m=1}^{N} \frac{R_n R_m}{R_n + R_m} \overline{PER}_n^{sd} P_m^{rd} P_n^{sd} \tag{23}$$

where

$$P_n^{sd} = \int_{\gamma_{1,n}}^{\gamma_{1,n+1}} p_{\gamma_1}(\gamma) d\gamma = F\left(\gamma_{1,n}, \gamma_{1,n+1}, \underline{c}^{(sd)}\right) \tag{24}$$

Here $\underline{c}^{(sd)} = \left(A^{sd}, k^{sd}, \bar{\gamma}_u^{sd}, \mu_s^{sd}, \sigma_s^{sd}\right)$ is a vector that contains the S-D channel parameters, and the function $F(\cdot,\cdot,\cdot)$ is defined as

$$F(x,y,\underline{z}) = (1-z_1)\left(Q_1(\sqrt{2z_2}, \sqrt{2xv}) - Q_1(\sqrt{2z_2}, \sqrt{2yv})\right) + (z_1)(\phi_t(-x) - \phi_t(-y)) \tag{25}$$

where $\underline{z} = (z_1, z_2, z_3, z_4, z_5)$, $v = (1 + z_2)/z_3$, $Q_1(\cdot,\cdot)$ is the first order Marcum $Q$-function [28], and the expression $\phi_t(s) = \int_0^\infty e^{-st} \frac{\xi}{\sqrt{2\pi}z_5 t} \exp\left\{\frac{-(10\log_{10}^t + z_4)^2}{2z_5^2}\right\} dt$, is the moment generating function of the random variable $t$ which is lognormally distributed with the mean $-z_4$ and variance $z_5^2$. The probability $P_m^{rd}$, is also computed by substituting the R-D channel parameters into (24). The average PER $\overline{PER}_n^{sd}, n = 1,2,\ldots,N$ in (23) can be obtained based on the equation (3) as follows

$$\overline{PER}_n^{sd} = \frac{G(\gamma_{1,n}, \gamma_{1,n+1}, \underline{c}^{(sd)})}{F(\gamma_{1,n}, \gamma_{1,n+1}, \underline{c}^{(sd)})} \tag{26}$$

where

$$G(x, y, \underline{z}) = (1 - z_1) \frac{a_n v}{g_n + v} \exp\left(\frac{-g_n z_2}{g_n + v}\right) \left[Q_1\left(\sqrt{\frac{2 z_2 v}{g_n + v}}, \sqrt{2x(g_n + v)}\right) - Q_1\left(\sqrt{\frac{2 z_2 v}{g_n + v}}, \sqrt{2y(g_n + v)}\right)\right]$$

$$+ z_1 \int_0^\infty \frac{a_n}{g_n w + 1} \left(e^{-(g_n + 1/w)x} - e^{-(g_n + 1/w)y}\right) \frac{\xi}{\sqrt{2\pi}z_5 w} \exp\left\{\frac{-(10\log_{10}^w - z_4)^2}{2z_5^2}\right\} dw \tag{27}$$

The average PER, $\overline{PER}_0^{sd}$, in the outage mode of S-D channel is also obtained based on the equations (1) and (3) as follows

$$\overline{PER}_0^{sd} = \frac{F(0, \Gamma_1, \underline{c}^{(sd)}) + G(\Gamma_1, \gamma_{1,1}, \underline{c}^{(sd)})}{F(0, \gamma_{1,1}, \underline{c}^{(sd)})} \tag{28}$$

*Proof:* Following the same approach presented in Appendix A and based on the proposed equations in (24)-(28), deriving the expression (23) is straightforward.

The next corollary also presents a closed form expression for the average PLR.

*Corollary 4:* The average system PLR for the considered LMSC system at $N_r = 1$, is given by

$$\overline{PLR} = \left(\sum_{n=0}^{N} \overline{PER}_n^{sd} P_n^{sd}\right) \left(\frac{\sum_{m=1}^{N} \overline{PER}_m^{rd} P_m^{rd}}{\sum_{m=1}^{N} P_m^{rd}}\right) \tag{29}$$

where the average PER, $\overline{PER}_m^{rd}, m = 1,2,\ldots,N$ is computed based on the R-D channel parameters.

*Proof:* Taking a similar approach as that presented in Appendix B, the proof is straightforward.

Despite the fact that the performance metrics for LMSC system are different from those in section 3.1, the proposed cross-layer design for this system has the same structure as in section 3.2, except that the step 4 is modified as follows.

To derive the equation (12) in the new scenario, we also consider the effect of data transmission in the S-D channel outage mode, over the PLR QoS constraint as follows

$$\left(P_{out}\overline{PER}_0^{sd} + P_{t,sd}(1 - P_{out})\right) P_{t,rd} = P_{loss} \tag{30}$$

where $P_{out} = \int_0^{\gamma_{1,1}} p_{\gamma_1}(\gamma) d\gamma = F(0, \gamma_{1,1}, \underline{c}^{(sd)})$.

Accordingly, we modify the target PER for the R-D channel, in step 4 of the design algorithm, based on the following equation

$$P_{t,rd} = \frac{P_{loss}}{P_{t,sd}(1-P_{out})+P_{out}\overline{PER}_0^{sd}} \tag{31}$$

As a final note, we refer to the fixed rate cooperative-ARQ as a promising scheme in the LMSC, when the required CSI for rate adaptation may not be available due to the rapid channel gain variations. Using the derivations in section 3.3, it is straightforward to develop an optimized fixed rate ARQ scheme for the LMSC system. The derivations are omitted here due to space limitations, but the results are presented and discussed in section 5.2.

## 5. Numerical Results

In this section, we provide numerical results to evaluate the performance of the proposed schemes. We denote the S-D, R-D, and S-R distances in Fig. 1 by $d_1$, $d_2$ and $d_3$, respectively. Assuming an identical noise variance $\sigma^2$ for all channels, the SNR of S-D, R-D and S-R channels are given by $\bar{\gamma}_1 = K_1 \bar{P} d_1^{-n}/\sigma^2$, $\bar{\gamma}_2 = K_2 \bar{P} d_2^{-n}/\sigma^2$ and $\bar{\gamma}_{sr} = K_3 \bar{P} d_3^{-n}/\sigma^2$, respectively, where $K_i d_i^{-n}, i = 1,2,3$ denote the path losses and the parameters $K_i, i = 1,2,3$ depend on the link parameters [29]. In our experiments, we normalize the aforementioned channel SNRs as $\bar{\gamma}_1 = \bar{P}$, $\bar{\gamma}_{sr} = \alpha \bar{P}$ and $\bar{\gamma}_2 = \lambda \bar{P}$, where, the parameters $\alpha$ and $\lambda$ are determined by large-scale path losses as $\lambda = K_2 d_2^{-n}/K_1 d_1^{-n}$ and $\alpha = K_3 d_3^{-n}/K_1 d_1^{-n}$.

For both source and relay nodes, we select five AMC modes adopted from the HYPERLAN/2 standard. Table II from [16] presents these AMC modes and the corresponding fitting parameters for a packet length $N_P = 1080$ bits. Naturally, one may consider other AMC modes in the presented framework. In all experiments, we consider a target PLR $P_{loss} = 0.001$. In the following, we first present the results for terrestrial links with Rayleigh fading model. We then evaluate packet communications over the LMSC system with a two-state Lutz's channel model.

### 5.1. Performance Analysis for Rayleigh Fading Channel

For this channel model, the S-D and R-D channel SNRs follow independent exponential distributions with the statistical average means $\bar{P}$, $\alpha \bar{P}$ and $\lambda \bar{P}$, respectively.

Fig. 2 depicts the average spectral efficiency versus the average SNR of S-D channel for different transmission schemes, where the S-D channel varies slowly as in [30]. In [30], the effect of rate adaptation in conjunction with a conventional ARQ protocol is examined, where the channel gain remains constant over transmission and possible retransmissions of a packet. The proposed analytical design framework may be used in this setting following two steps: (1) First, we derive the corresponding performance metrics including spectral efficiency and PLR, using the analysis presented in Appendices A and B and by considering $\gamma_1 = \gamma_2$, as follows

$$\eta = \sum_{n=1}^{N} R_n (1 - \tfrac{1}{2}\overline{PER_n^{sd}}) P_n^{sd}$$

$$\overline{PLR} = \frac{\sum_{n=1}^{N} \overline{PLR}_n P_n^{sd}}{\sum_{n=1}^{N} P_n^{sd}}$$

where $\overline{PLR}_n = \frac{1}{P_n^{sd}} \int_{\gamma_{1,n}}^{\gamma_{1,n+1}} PER_n^2(\gamma)\, p_{\gamma_1}(\gamma) d\gamma$. (2) We then design the AMC and obtain the mode switching levels from $\overline{PLR}_n = P_{loss}$ following the approach proposed in [19]. This scheme is referred to as conventional ARQ-AMC in Fig. 2. We also use an AMC only scheme [19] on the S-D link as a benchmark for performance comparison.

As evident in Fig. 2, the proposed joint cooperative ARQ-AMC scheme considerably improves the spectral efficiency when compared to the two other schemes. This in turn signifies the role of retransmission by the relay, which relieves the stringent error performance requirements on the S-D link. Fig. 2, also shows that using the proposed algorithm to optimize the target PER over S-D and R-D links, improves the system spectral efficiency in comparison with a system that uses equal target PERs for these links ($P_{t,rd} = P_{t,sd}$). From this figure, we also observe that a better position of the relay node, which results in a higher R-D channel SNR, increases the system spectral efficiency ($\lambda$=10 vs. $\lambda$=0).

In Fig. 3, we plot the spectral efficiency of the joint cooperative ARQ-AMC scheme for different S-R channel qualities (SNRs). As evident, a better S-R channel quality, leads to a higher spectral efficiency ($\alpha$=10 vs. $\alpha$=0), and when the S-R channel is of poor quality ($\alpha$=-10), the performance of the proposed scheme is very close to that of a direct-transmission scheme with AMC alone. We also observe that when the S-R channel SNR exceeds a threshold (here, $\alpha \geq 10$ dB), the S-R channel can be considered as error free. Note that for $\alpha = 0$ dB, the search region in the algorithm of section 3.2 is bounded to guarantee that $P_{t,rd} > 0$, as a result the corresponding spectral efficiency curve in Fig. 3 is not perfectly smooth.

In Fig. 4, we compare the average spectral efficiency of adaptive rate and optimized fixed rate cooperative ARQ schemes. It is observed that the adaptive rate cooperative ARQ scheme considerably outperforms the optimized fixed rate schemes. Obviously, this performance gain is attributed to using instantaneous per frame channel CSI at the source and relay nodes, compared to the case where only channel statistics are used. This observation signifies the role of exploiting channel CSI jointly with cooperative ARQ in a relay channel. Specifically, this performance gain is noticeable for low S-D channel SNRs, where a fixed rate scheme cannot satisfy the PLR QoS constraint and results in a poor spectral efficiency. In Fig. 4, we also see that selecting different transmission rates for S and R nodes improves the spectral efficiency, when compared to the scheme that uses equal rates at these nodes.

As specified in Corollaries 1 and 2, the proposed scheme for $\alpha = \infty$ and $\lambda = 1$ reduces to an AMC scheme with conventional ARQ. As stated, in the proposed formulation different optimized target PERs are considered for the transmission and possible retransmission of a packet. In [16] a different approach is proposed, which considers identical target PERs. As depicted in Fig. 5, the proposed scheme outperforms the scheme of [16], especially for smaller average SNRs.

**5.2. Performance Analysis for LMSC System**

In the following numerical results, we use the channel parameters of city and highway environments for S-D and R-D channels, respectively as presented in [27]. In this setting, the relay terminal is assumed to be a high altitude platform station which can provide different R-D channel qualities. Table I shows the channel parameters based on the experimental measurements of [27]. We also select the transmission rate in the S-D channel outage mode as $R_0 = R_1$.

Fig. 6 shows the average spectral efficiency of different adaptive rate cooperative ARQ schemes in LMSC system. As evident from this figure considering transmission in the outage mode of S-D channel, the proposed joint cooperative ARQ-AMC scheme dramatically increases the system spectral efficiency when the S-D link has a low average SNR. In fact, in this setting, the outage mode of LMSC system has a high probability. Moreover, the proposed scheme outperforms the conventional ARQ scheme when the S-D channel is subject to a high temporal correlation. These substantial spectral efficiency gains signify the role of relay retransmission for blockage mitigation in LMSC systems.

In Fig. 7, we plot the spectral efficiency of the adaptive rate and fixed rate cooperative ARQ schemes. As evident, combining AMC with cooperative ARQ provides much higher spectral efficiency gain when compared to the fixed-rate cooperative ARQ thanks to the use of CSI at the source and relay transmitters. This figure also shows that a scheme that considers independent and possibly different transmission rates for S and R nodes outperforms the scenario where the S and R nodes are constrained to choose equal transmission rates.

Comparing the results in sections 5.1 and 5.2, illustrates that the proposed scheme yields a higher performance gain in LMSC systems. This is because, in this scenario the channel blockage effect is compensated effectively using a cooperative ARQ protocol instead of a conventional ARQ.

**6. Conclusions**

In this paper, we developed a cross-layer approach to jointly design AMC at the physical layer and cooperative ARQ at the data link layer to enhance the system performance for data packet transmission over block fading relay channels. The proposed scheme maximizes the system spectral efficiency subject to a prescribed PLR constraint for delay constrained packet services. We have shown that the presented framework can be well fitted to applications such as LMSC, where the channel blockage effects severely degrades the performance of conventional ARQ schemes. Numerical results indicate a considerable spectral efficiency gain when compared to systems such as AMC at the physical layer alone, fixed (optimized) rate cooperative ARQ, fixed equal rate cooperative ARQ, and joint conventional ARQ-AMC scheme. This in turn validates the efficiency of the proposed cross-layer approach for QoS provisioning in wireless relay packet networks.

Currently we are developing similar cross-layer approaches for adaptive transmission policy design in wireless relay networks with bursty and delay-sensitive packet traffic.

## Appendix A

For an adaptive rate system that utilizes Nyquist pulses, the spectral efficiency is the average number of information bits per symbol [2]. Let us now consider a packet based system where each packet contains a fixed number of $N_P$ bits, transmitted using $L$ symbols. In general, the average number of transmitted bits per symbol is given by

$$\eta = E\left[\frac{N_P}{L}\right] \tag{32}$$

where $E[.]$ denotes the expectation operator. For an adaptive rate cooperative $N_r$-truncated ARQ protocol, the relay node retransmits the erroneously received packets, until it is received correctly or a maximum allowable number of transmissions is reached. Therefore, each packet data, in general, encounters a vector of channel SNR realizations denoted by $\underline{\gamma} = (\gamma_1, \gamma_2^1, \ldots, \gamma_2^M)$. Here, the random variable $\gamma_1$ denotes the SNR of S-D channel and $\gamma_2^l$, $l = 1,2,\ldots,M$ denotes the SNR of R-D channel for possible $M$ retransmissions of a packet. The random variable $M \in \{1,2,\ldots,N_r\}$ depends on the channel noise. Let $R_n$ and $R_{m_l}$, $l = 1,\ldots,M; n, m_l \in \{1,\ldots,N\}$ be random variables, which show the selected rates by the source and relay nodes based on the channel SNRs $\gamma_1$ and $\gamma_2^l$, respectively. Then, the number of transmitted symbols per packet for channel SNR $\gamma_1$ is $L_s(0) = N_P/R_n$, and for channel SNR $\gamma_2^l$ is $L_s(l) = N_P/R_{m_l}$, $l = 1,\ldots,M$. Therefore, the instantaneous spectral efficiency is given by

$$\eta(\underline{\gamma}, M) = 1/L(n, \underline{m}^{(M)}), \quad M = 0,1,\ldots,N_r \tag{33}$$

where $L(n, \underline{m}^{(M)}) = \sum_{l=0}^{M} L_s(l)/N_P = 1/R_n + \sum_{l=1}^{M} 1/R_{m_l}$ and $\underline{m}^{(M)} = (m_1, m_2, \ldots, m_M)$. The random variable $M$ can be statistically described as follows

$$M = \begin{cases} 0, & (T_{sd}:s) \cup (T_{sd}:f \cap T_{sr}:f) \\ k, & (T_{sd}:f \cap T_{sr}:s) \cap (T_{rd}^1:f \cap \cdots \cap T_{rd}^{k-1}:f) \cap T_{rd}^k:s, \\ & k = 1,2,\ldots,N_r - 1 \\ N_r, & (T_{sd}:f \cap T_{sr}:s) \cap (T_{rd}^1:f \cap \cdots \cap T_{rd}^{N_r-1}:f) \end{cases} \tag{34}$$

where $T_{sd} \in \{s, f\}$ and $T_{rd}^k \in \{s, f\}, k \in \{1, 2, \ldots, N_r\}$ are the events indicating the success ($s$) or failure ($f$) of the transmission over the S-D channel and $k$'th retransmission over the R-D channel, respectively. Also, we have

$$\Pr(T_{sd}: f) = PER_n(\gamma_1)$$
$$\Pr(T_{rd}^k: f) = PER_{m_k}(\gamma_2^k), \quad k = 1, \ldots, N_r$$
$$\Pr(T_{sr}: f) = \varepsilon_n$$

In general, the average spectral efficiency of the proposed joint cooperative truncated ARQ-AMC scheme can be written as

$$\eta = E_{\underline{\gamma}} E_M \{\eta(\underline{\gamma}, M) | \underline{\gamma}\} \tag{35}$$

Averaging with respect to the random variable $M$, the inner expectation in (35) is given by

$$\eta(\underline{\gamma}) = E_M \{\eta(\underline{\gamma}, M) | \underline{\gamma}\}$$
$$= \sum_{l=0}^{N_r} \frac{1}{L(n, \underline{m}^{(l)})} \Pr(M = l) \tag{36}$$

As a special case for $N_r = 0$, we have $\Pr(M = 0) = 1$ and (36) is reduced to

$$\eta = \sum_{n=1}^{N} R_n P_n^{sd} \tag{37}$$

where $P_n^{sd} = \int_{\gamma_{1,n}}^{\gamma_{1,n+1}} p_{\gamma_1}(\gamma) d\gamma$. The equation (37) is the spectral efficiency of AMC-only scheme [2]. For the general case of $N_r \geq 1$, using (34) and (36), we have

$$\eta(\underline{\gamma}) = \frac{1}{L(n, \underline{m}^{(0)})} \left(1 - (1 - \varepsilon_n) PER_n(\gamma_1)\right)$$
$$+ \sum_{l=1}^{N_r - 1} \frac{(1 - \varepsilon_n)}{L(n, \underline{m}^{(l)})} PER_n(\gamma_1) \prod_{k=1}^{l-1} PER_{m_k}(\gamma_2^k) \left(1 - PER_{m_l}(\gamma_2^l)\right) \tag{38}$$
$$+ \frac{1}{L(n, \underline{m}^{(N_r)})} (1 - \varepsilon_n) PER_n(\gamma_1) \prod_{k=1}^{N_r - 1} PER_{m_k}(\gamma_2^k)$$

In deriving the equation (38), we use the fact that the channel gains in transmission and possible retransmissions of a packet are independent, which is a consequence of block-fading assumption. The average spectral efficiency in (35) is given by

$$\eta = E_{\underline{\gamma}} \{\eta(\underline{\gamma})\} = \sum_{n=1}^{N} \int_{\gamma_{1,n}}^{\gamma_{1,n+1}} \frac{1}{L(n, \underline{m}^{(0)})} \left(1 - (1 - \varepsilon_n) PER_n(\gamma_1)\right) p_{\gamma_1}(\gamma_1) d\gamma_1$$
$$+ \sum_{l=1}^{N_r - 1} \sum_{n=1}^{N} \sum_{m_1=1}^{N} \cdots \sum_{m_l=1}^{N} \int_{\gamma_{1,n}}^{\gamma_{1,n+1}} \int_{\gamma_{2,m_1}}^{\gamma_{2,m_1+1}} \cdots \int_{\gamma_{2,m_l}}^{\gamma_{2,m_l+1}} (1 - \varepsilon_n)$$
$$\times \left\{ \frac{PER_n(\gamma_1) \prod_{k=1}^{l-1} PER_{m_k}(\gamma_2^k) \left(1 - PER_{m_l}(\gamma_2^l)\right)}{L(n, \underline{m}^{(l)})} \right\} p_{\gamma_1}(\gamma_1) \prod_{k=1}^{l} p_{\gamma_2^k}(\gamma_2^k) d\gamma_2^k d\gamma_1 \tag{39}$$
$$+ \sum_{n=1}^{N} \sum_{m_1}^{N} \cdots \sum_{m_{N_r}=1}^{N} \int_{\gamma_{1,n}}^{\gamma_{1,n+1}} \int_{\gamma_{2,m_1}}^{\gamma_{2,m_1+1}} \cdots \int_{\gamma_{2,m_{N_r}}}^{\gamma_{2,m_{N_r}+1}} (1 - \varepsilon_n)$$

$$\times \left\{ PER_n(\gamma_1) \frac{\prod_{k=1}^{N_r-1} PER_{m_k}(\gamma_2^k)}{L(n,\underline{m}^{(N_r)})} \right\} p_{\gamma_1}(\gamma_1) \prod_{k=1}^{N_r} p_{\gamma_2^k}(\gamma_2^k) d\gamma_2^k d\gamma_1$$

By defining

$$P_n^{sd} := \int_{\gamma_{1,n}}^{\gamma_{1,n+1}} p_{\gamma_1}(\gamma) d\gamma \tag{40}$$

$$P_m^{rd} := \int_{\gamma_{2,m}}^{\gamma_{2,m+1}} p_{\gamma_2}(\gamma) d\gamma \tag{41}$$

$$\overline{PER}_n^{sd} := \frac{1}{P_n^{sd}} \int_{\gamma_{1,n}}^{\gamma_{1,n+1}} PER_n(\gamma) p_{\gamma_1}(\gamma) d\gamma \tag{42}$$

$$\overline{PER}_m^{rd} := \frac{1}{P_m^{rd}} \int_{\gamma_{2,m}}^{\gamma_{2,m+1}} PER_m(\gamma) p_{\gamma_2}(\gamma) d\gamma \tag{43}$$

and after following a few steps, we obtain

$$\eta = \sum_{n=1}^{N} \frac{(1-(1-\varepsilon_n)\overline{PER}_n^{sd})}{L(n,\underline{m}^{(0)})} P_n^{sd} + \sum_{l=1}^{N_r-1} \sum_{n=1}^{N} \sum_{m_1=1}^{N} \cdots \sum_{m_l=1}^{N} \frac{(1-\varepsilon_n)}{L(n,\underline{m}^{(l)})}$$

$$\times \overline{PER}_n^{sd} P_n^{sd} \prod_{k=1}^{l-1} \overline{PER}_{m_k}^{rd} P_{m_k}^{rd} \left(1 - \overline{PER}_{m_l}^{rd}\right) P_{m_l}^{rd}$$

$$+ \sum_{n=1}^{N} \sum_{m_1}^{N} \cdots \sum_{m_{N_r}=1}^{N} \frac{(1-\varepsilon_n)\overline{PER}_n^{sd} P_n^{sd} P_{m_{N_r}}^{rd}}{L(n,\underline{m}^{(N_r)})} \prod_{k=1}^{N_r-1} \overline{PER}_{m_k}^{rd} P_{m_k}^{rd} \tag{44}$$

## Appendix B

Given the instantaneous SNRs $\gamma_1$, $\gamma_2$, and a fixed SNR $\gamma_{sr}$ for the S-D, R-D, and S-R channels, respectively, using the total probability theorem, the average PLR of the proposed scheme is given by

$$\overline{PLR}(\gamma_{sr}) = \iint_0^\infty \Pr(\text{Loss of packet}|\gamma_1,\gamma_2) p_{\gamma_1,\gamma_2}(\gamma_1,\gamma_2) d\gamma_1 d\gamma_2 \tag{45}$$

Where

$$\Pr(\text{Loss of packet}|\gamma_1,\gamma_2) = \Pr(T_{sd}:f,T_{sr}:f|\gamma_1,\gamma_2) + \Pr(T_{sd}:f,T_{sr}:s,T_{rd}:f|\gamma_1,\gamma_2) \tag{46}$$

Since the channels noise and channel SNRs $\gamma_1$ and $\gamma_2$ are independent, the equation (46) reduces to

$$\Pr(\text{Loss of packet}|\gamma_1,\gamma_2) = \Pr(T_{sd}:f|\gamma_1)\Pr(T_{sr}:f|\gamma_1)$$
$$+ \Pr(T_{sd}:f|\gamma_1)\Pr(T_{rd}:f|\gamma_2)\Pr(T_{sr}:s|\gamma_1)$$
$$= PER_n(\gamma_1)\varepsilon_n + (1-\varepsilon_n)PER_n(\gamma_1)PER_m(\gamma_2) \tag{47}$$

where $T_{sd}, T_{sr}, T_{rd}, f$, and $s$ are described in Appendix A, and $n, m$ are the AMC modes used for S-D and R-D links respectively. In the scenario under consideration, in the outage modes of S-D (*i.e.* $\gamma_1 < \gamma_{1,1}$) and R-D links (*i.e.* $\gamma_2 < \gamma_{2,1}$), no data is transmitted by the S and R nodes. Therefore, using (45) the average PLR can be calculated as

$$\overline{PLR}(\gamma_{sr}) = \iint_0^\infty \Big[\Pr(T_{sd}:f, T_{sr}:f|\gamma_1,\gamma_2,\gamma_1>\gamma_{1,1})$$
$$+ \Pr(T_{sd}:f, T_{sr}:s, T_{rd}:f|\gamma_1,\gamma_2,\gamma_1>\gamma_{1,1},\gamma_2>\gamma_{2,1})\Big] p_{\gamma_1}(\gamma_1) p_{\gamma_2}(\gamma_2) d\gamma_1 d\gamma_2 \qquad (48)$$
$$= \frac{1}{\Pr(\gamma_1>\gamma_{1,1})} \int_{\gamma_{1,1}}^\infty \Pr(T_{sd}:f, T_{sr}:f|\gamma_1) p_{\gamma_1}(\gamma_1) d\gamma_1$$
$$+ \frac{1}{\Pr(\gamma_1>\gamma_{1,1},\gamma_2>\gamma_{2,1})} \int_{\gamma_{2,1}}^\infty \int_{\gamma_{1,1}}^\infty \Pr(T_{sd}:f, T_{sr}:s, T_{rd}:f|\gamma_1,\gamma_2) p_{\gamma_1}(\gamma_1) p_{\gamma_2}(\gamma_2) d\gamma_1 d\gamma_2$$

Substituting the equation (47) in (48), we can obtain

$$\overline{PLR}(\gamma_{sr}) = \frac{\sum_{n=1}^N \varepsilon_n \int_{\gamma_{1,n}}^{\gamma_{1,n+1}} PER_n(\gamma_1) p_{\gamma_1}(\gamma_1) d\gamma_1}{\sum_{n=1}^N \int_{\gamma_{1,n}}^{\gamma_{1,n+1}} p_{\gamma_1}(\gamma_1) d\gamma_1} + \left(\frac{\sum_{n=1}^N (1-\varepsilon_n) \int_{\gamma_{1,n}}^{\gamma_{1,n+1}} PER_n(\gamma_1) p_{\gamma_1}(\gamma_1) d\gamma_1}{\sum_{n=1}^N \int_{\gamma_{1,n}}^{\gamma_{1,n+1}} p_{\gamma_1}(\gamma_1) d\gamma_1}\right)$$
$$\times \left(\frac{\sum_{m=1}^N \int_{\gamma_{2,m}}^{\gamma_{2,m+1}} PER_m(\gamma_2) p_{\gamma_2}(\gamma_2) d\gamma_2}{\sum_{m=1}^N \int_{\gamma_{2,m}}^{\gamma_{2,m+1}} p_{\gamma_2}(\gamma_2) d\gamma_2}\right) \qquad (49)$$

Using the equations (40) to (43) and following a few steps, the PLR in equation (49) is expressed as

$$\overline{PLR}(\gamma_{sr}) = \left(\frac{\sum_{n=1}^N \overline{PER_n^{sd}} P_n^{sd}}{\sum_{n=1}^N P_n^{sd}}\right)\left(\frac{\sum_{m=1}^N \overline{PER_m^{rd}} P_m^{rd}}{\sum_{m=1}^N P_m^{rd}}\right) + \left(\frac{\sum_{n=1}^N \varepsilon_n \overline{PER_n^{sd}} P_n^{sd}}{\sum_{n=1}^N P_n^{sd}}\right) \times \left(1 - \frac{\sum_{m=1}^N \overline{PER_m^{rd}} P_m^{rd}}{\sum_{m=1}^N P_m^{rd}}\right) \qquad (50)$$

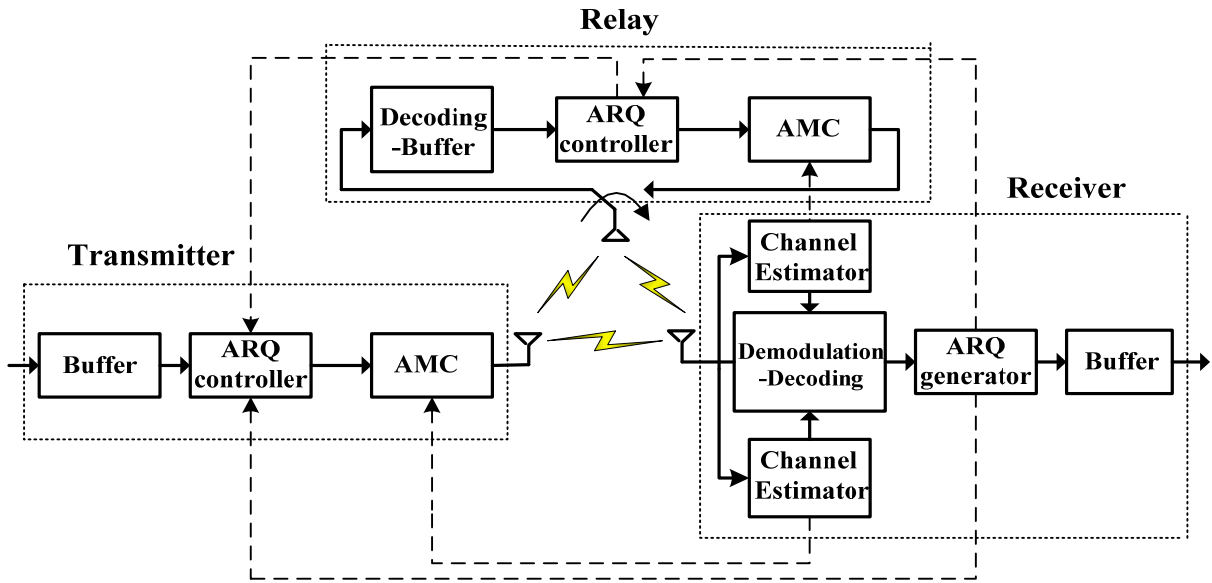

Fig. 1 System model.

TABLE I
The S-D and R-D channel parameters in LMSC system when the power of unfaded satellite link is normalized to unity [27].

| Channel | A | $k$ (dB) | $\mu_s$ (dB) | $\sigma_s$ (dB) |
|---|---|---|---|---|
| S-D | 0.89 | 3.9 | -11.5 | 2.0 |
| R-D | 0.24 | 10.2 | -8.9 | 5.1 |

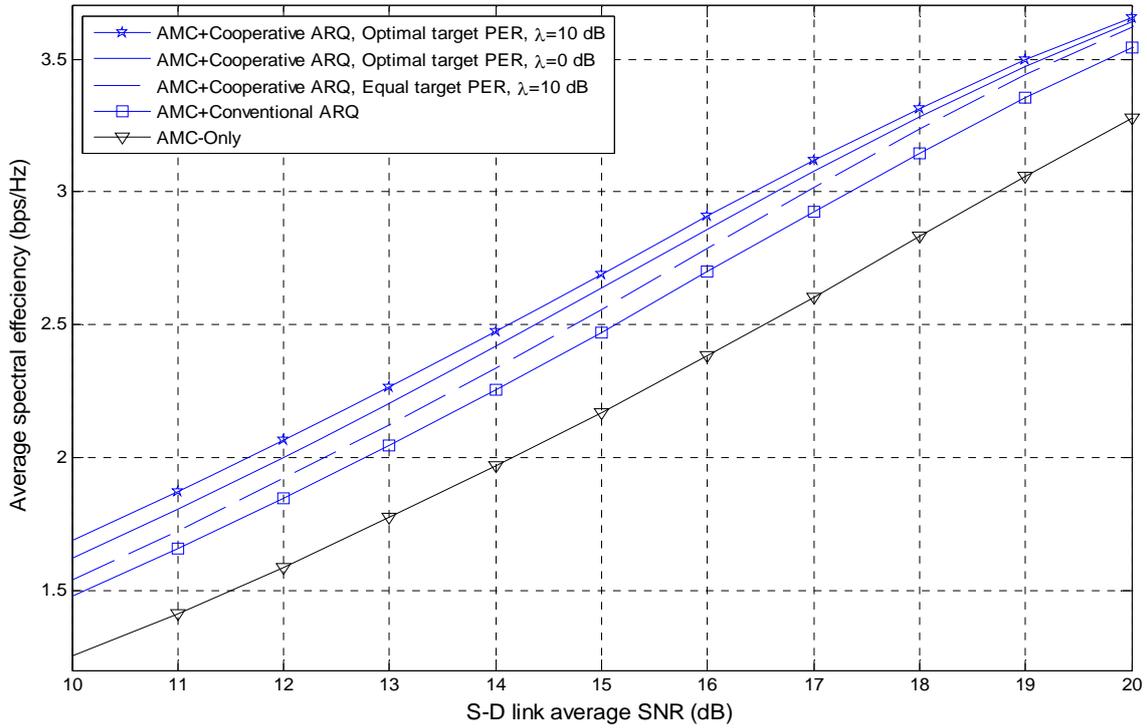

Fig. 2. Spectral efficiency vs. the average SNR of S-D channel for joint cooperative ARQ-AMC and AMC with/without conventional ARQ schemes, $\alpha$=10 dB. The S-D channel is assumed to be slowly varying [30].

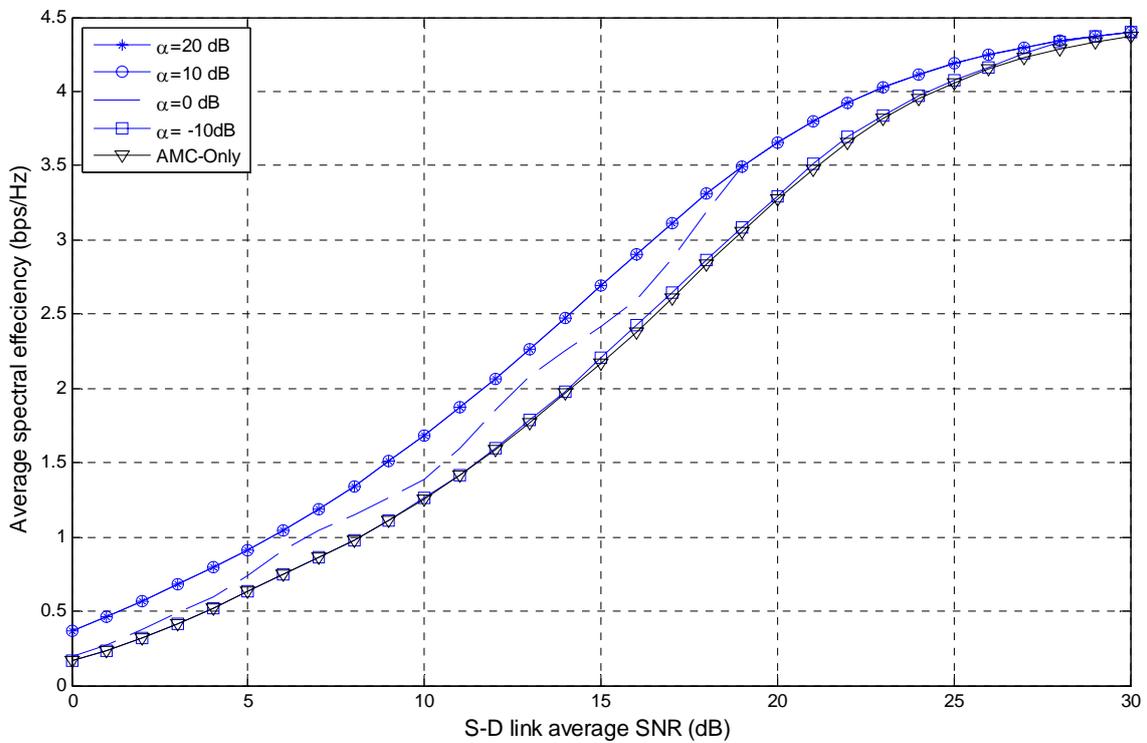

Fig. 3. Spectral efficiency vs. the average SNR of S-D channel for joint cooperative ARQ-AMC scheme with different S-R channel SNRs, $\lambda$=10 dB.

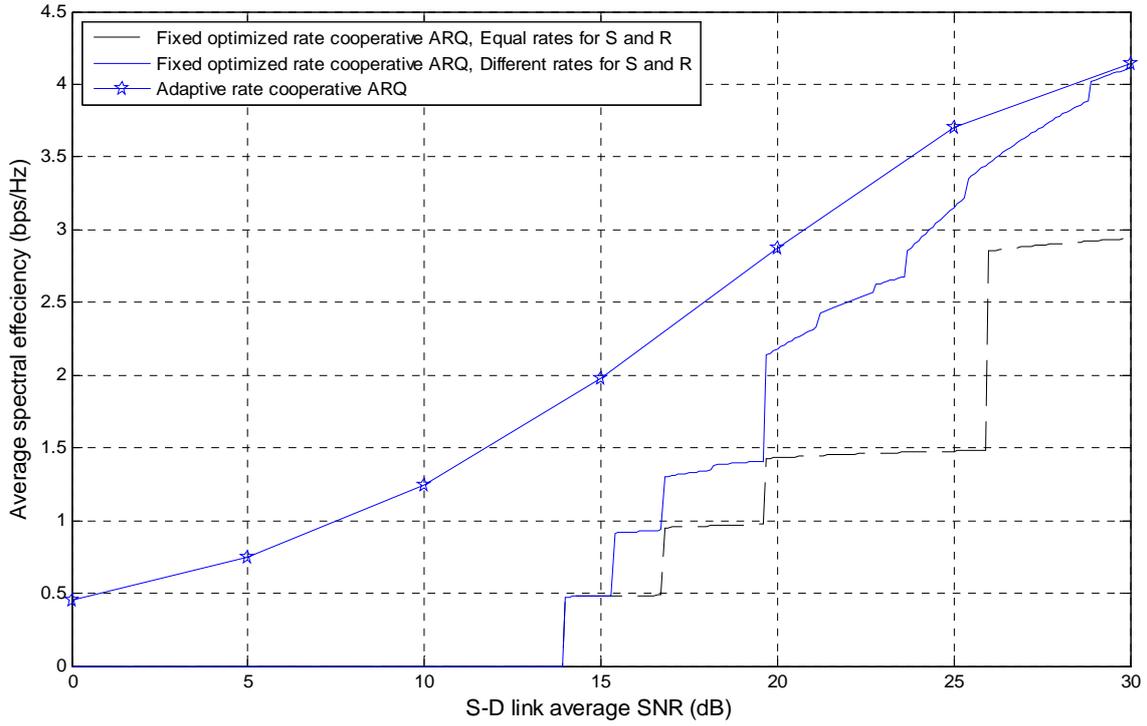

Fig. 4. Spectral efficiency vs. the average SNR of S-D channel for adaptive rate cooperative ARQ and fixed rate cooperative ARQ schemes, $\alpha$=10 dB, $\lambda$=10 dB.

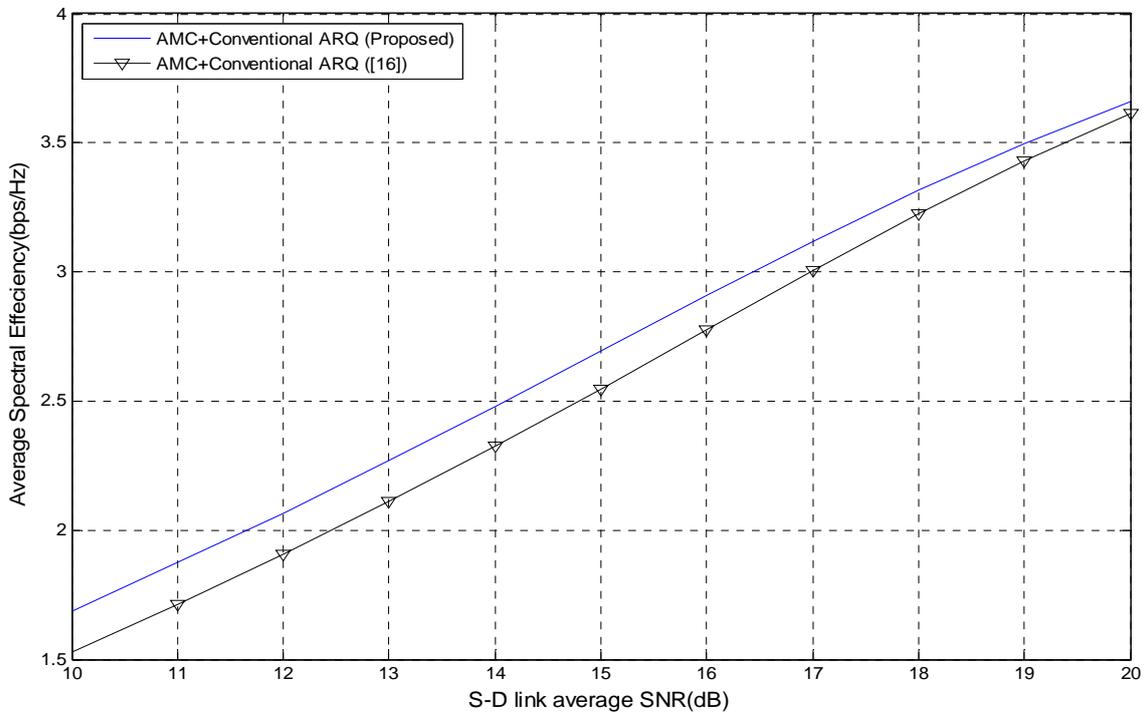

Fig. 5. Spectral efficiency vs. the average SNR of S-D channel for the proposed conventional ARQ scheme and that in [16].

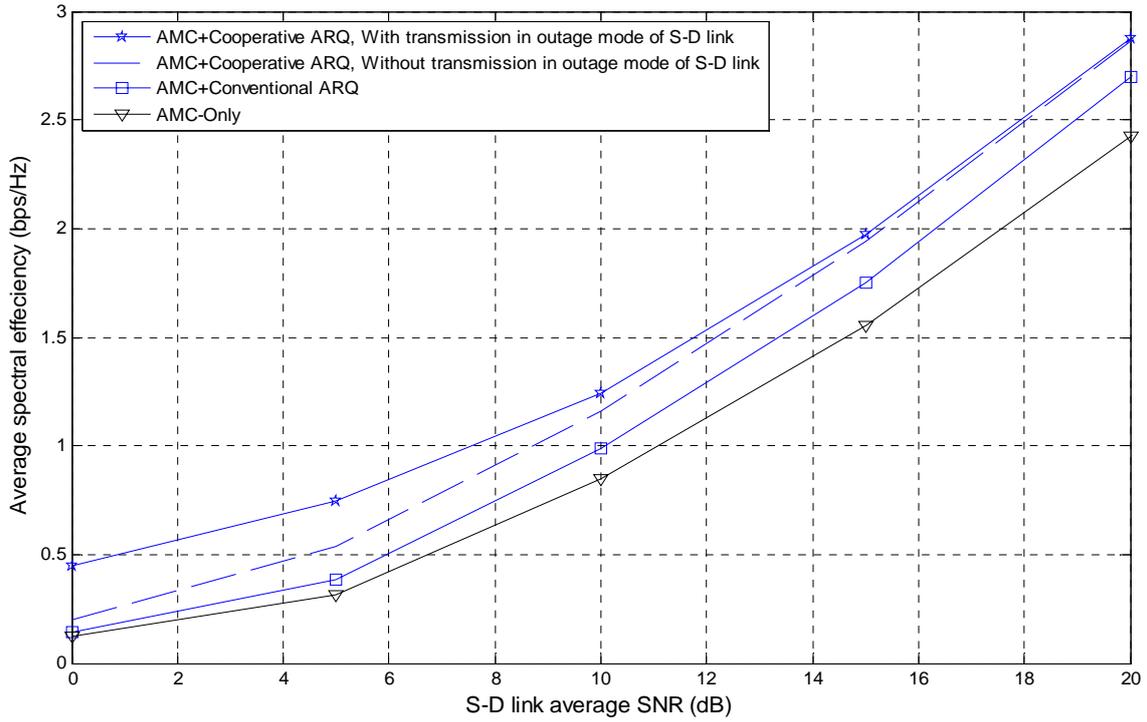

Fig. 6. Spectral efficiency vs. the average SNR of S-D channel for different adaptive rate transmission schemes, $\lambda=10$ dB. The S-D channel is assumed to be slowly varying [30].

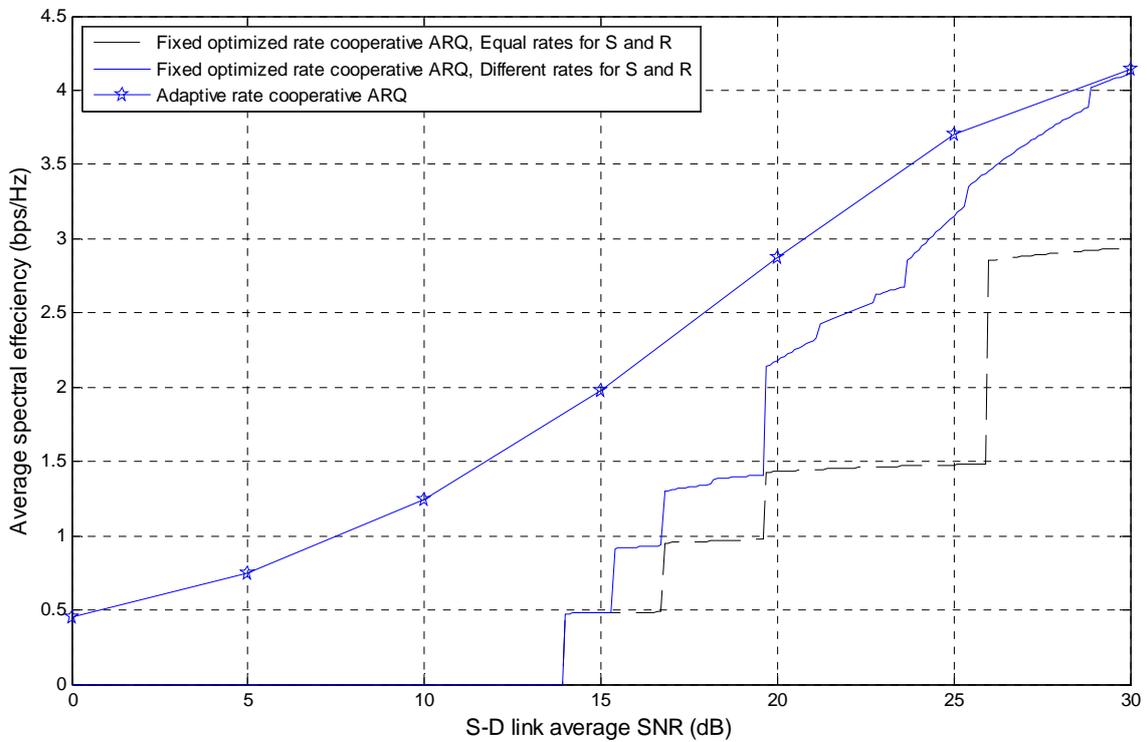

Fig. 7. Spectral efficiency vs. the average SNR of S-D channel for adaptive rate and fixed rate cooperative ARQ schemes, $\lambda=10$ dB.